\documentclass{elsart}
\usepackage{latexsym,amssymb,amsmath,amsfonts,epsfig}

\makeatletter \@addtoreset{equation}{section} \makeatother
\renewcommand{\theequation}{\thesection.\arabic{equation}}
\newcommand {\lhs}       {left-hand  side}
\newcommand {\rhs}       {right-hand side}

\newcommand {\CS}        {Chern--Simons}

\newcommand {\bdi}{\begin{eqnarray*}}
\newcommand {\edi}{\end{eqnarray*}}
\newcommand {\no}{\nonumber}
\newcommand {\gsim}{\mathrel{\hbox{\rlap{\lower.55ex \hbox {$\sim$}}
            \kern-.3em \raise.4ex \hbox{$>$}}}}
\newcommand {\lsim}{\mathrel{\hbox{\rlap{\lower.55ex \hbox {$\sim$}}
            \kern-.3em \raise.4ex \hbox{$<$}}}}

\def\id{\makebox[0.6ex][l]{$1$}{\rm l}}

\def\1{\id}                             

\begin{document}
\noindent  Nucl. Phys. B 657 (2003) 214 
\hfill KA--TP--20--2002\vspace*{1cm}\runauthor{Adam and Klinkhamer}

\begin{frontmatter}
\title{Photon decay in a CPT-violating extension of
quantum electrodynamics}

\author[ITP,WPI]{C.~Adam},
\ead{adam@particle.uni-karlsruhe.de}
\author[ITP]{F.R.~Klinkhamer}
\ead{\\frans.klinkhamer@physik.uni-karlsruhe.de}

\address[ITP]{Institut f\"ur Theoretische Physik, Universit\"at
Karlsruhe, 76128 Karlsruhe,\newline
Germany}
\address[WPI]{Wolfgang Pauli Institute, c/o Institut f\"ur Mathematik,
Universit\"at Wien,\newline
1090 Wien, Austria\thanksref{note}}
\thanks[note]{present address}

\begin{abstract}
We consider the process of photon decay in quantum electrodynamics
with a CPT-violating Chern--Simons-like term added to the action.
For a simplified model with only the quadratic Maxwell and
Chern--Simons-like terms and the quartic Euler--Heisen\-berg term, we
obtain a nonvanishing probability for the decay of a particular
photon state into three others.
\end{abstract}

\end{frontmatter}

\section{Introduction}

The propagation of light in Maxwell theory with an extra
CPT-violating Lorentz-noninvariant \CS-like term in the action has
been studied both classically \cite{CFJ90} and quantum
mechanically \cite{AK01}. The Maxwell term and the \CS-like term
are quadratic in the photon field. These terms combined lead to
\emph{birefringence}, even in empty space. Additional
(Lorentz-invariant) higher-order photonic terms in the action
could perhaps produce other effects such as the \emph{decay} of
photons. (See also Ref.~\cite{KL01} for a discussion of the decay
of a massive Dirac fermion in theories with spontaneous Lorentz
and CPT violation.)

The purpose of this paper is to study the process of photon decay in detail,
expanding on the brief remark in Section 6 of Ref.~ \cite{AK01}.
In order to concentrate on the essentials, we keep
the theory as simple as possible. In fact, the theory we start
from is just quantum electrodynamics---the theory of photons, electrons and
positrons (see, e.g., Refs.~\cite{HP30,H36}).
It is, of course, known that the quantum effects of the electron-positron
field lead to a quartic coupling of the photons (giving, e.g.,
light-by-light scattering), which is described by the quartic part
of the so-called Euler--Heisenberg Lagrangian \cite{HE36}.

The photon model considered in this article has only the Maxwell and quartic
Euler--Heisenberg terms, together with a hypothetical spacelike \CS-like term
which breaks Lorentz and CPT invariance. For the moment, the precise origin of
this \CS-like term can be left open, but at least one possible mechanism
has been identified \cite{K00,KS02} (see Ref.~\cite{K01} for a review).
A more extensive discussion of the possible origin and
consequences of the \CS-like term can be found in Ref.~ \cite{AK01},
which also contains further references.

The outline of our paper is as follows. The model is
presented in Section 2, together with some basic facts on the
polarization states of the ``photons.''  The Lorentz noninvariance
of the model allows for photon decay and the relevant kinematics is
discussed briefly in Section 3. (Some technical details are
relegated to Appendix \ref{Energyinequalities}). The matrix element
for a particular decay channel is then calculated in Section 4
and the corresponding partial decay width for a photon at rest is
given in Section 5. (Numerical results for the phase space integral
are presented in Appendix \ref{Numericalresults}). For this particular
case, there is, in principle, one other decay channel available,
but it does not contribute as shown in Section 6 with details
relegated to Appendix \ref{Analyticresult}. The total decay rate
of the particular photon state at rest is discussed in Section 7.
The corresponding mean life time is expected to be large but
finite.

\section{Model}

The Lagrangian density of the model considered in this paper is
\begin{eqnarray} \label{calL}
\mathcal{L}=\mathcal{L}_0 +\mathcal{L}_1\,,
\end{eqnarray}
where the free part $\mathcal{L}_0$ consists of the usual Maxwell term \cite{HP30,H36}
and an additional \CS-like term \cite{CFJ90,AK01},
\begin{eqnarray}  \label{calL0}
\mathcal{L}_0 = -\frac{1}{4}F_{\mu\nu}F^{\mu\nu} +
\frac{1}{4}\, m \,\epsilon_{\mu\nu\rho\sigma}\, \eta^\mu A^\nu F^{\rho\sigma}\,.
\end{eqnarray}
Our conventions are $(g_{\mu\nu})=\mathrm{diag}(1,-1,-1,-1)$ and
$\epsilon_{0123}=1$, together with $\hbar$ $=$ $c$ $=$ $1$.

The \CS-like term in Eq.~(\ref{calL0}) has a mass parameter $m$.
(Ex\-per\-i\-men\-tal\-ly, there are tight constraints \cite{CFJ90,WPC97}
on this mass,
$m \lsim 10^{-33}\;\mathrm{eV}$, as will be discussed further in Section 7.)
The \CS-like term contains, in addition, a purely spatial ``four-vector''
\begin{eqnarray}  \label{etamu}
(\eta^\mu) = \left(0,\hat \eta^1 , \hat \eta^2 , \hat \eta^3 \right) \,,
\end{eqnarray}
in terms of a ``three-vector'' $\hat \eta$ of unit length, $|\hat\eta|^{\, 2} =1$.
The parameters $\eta^\mu$ are fixed once and for all, hence the quotation marks.
In fact, the condition of having fixed parameters (coupling constants) $\eta^\mu$
in the Lagrangian $\mathcal{L}_0$ effectively selects a class of
preferred inertial frames; cf. Refs.~\cite{CFJ90,AK01,KL01}.

In the following, we will also write the Maxwell field strength
$F_{\mu\nu}$ $\equiv$ $\partial_\mu A_\nu$ $-$ $\partial_\nu A_\mu$
in terms of the electric and magnetic fields,
$E_k \equiv F_{k0}$ and $B_k \equiv \epsilon_{klm} F_{lm}/2$
with $\epsilon_{123}=1$. If the gauge needs to be fixed,
we will use the radiation (Coulomb) gauge \cite{HP30,H36,W96},
\begin{eqnarray} \label{Coulomb}
\vec \partial \cdot \vec A=0 \,.
\end{eqnarray}
Further details on  the photon propagation and the microcausality of
the free theory (\ref{calL0})--(\ref{Coulomb}) can be found in
Ref.~\cite{AK01}.

The interaction term $\mathcal{L}_1$ in Eq.~(\ref{calL}) is taken to
be the quartic Euler--Heisenberg Lagrangian \cite{HE36}
\begin{eqnarray}   \label{calL1}
\mathcal{L}_1&=&
     \frac{2\,\alpha^2}{45\, M^4}\;
     \left[ \left(\, \frac{1}{2}\,F_{\mu\nu}F^{\mu\nu} \right)^2 +
         7\,\left(\, \frac{1}{8}\,\epsilon^{\mu\nu\rho\sigma}
                     F_{\mu\nu} F_{\rho\sigma}\right)^2
     \,\right]  \no\\
&=&  \frac{2\,\alpha^2}{45\, M^4}\;
     \left[ \left( \,|\vec E\,|^{\,2} - |\vec B\,|^{\,2} \,\right)^2 +
     7\,\left(\vec E\cdot\vec B\,\right)^2 \,\right]\,,
\end{eqnarray}
where $M$ is a mass-scale and $\alpha^2$ a dimensionless coupling constant.
For quantum electrodynamics with electron mass $M \approx 511\;\mathrm{keV}$
and fine-structure constant $\alpha \approx 1/137$,
precisely this term appears in the one-loop effective gauge field
action (the corresponding Feynman diagrams have a single electron loop with four
external photon lines). The quartic Euler--Heisenberg term in the effective
action is relevant for photon energies much less than $M$, which will be
the case for the process discussed in the present paper (with photon
energies less or equal to $m$). For further details on this effective action,
see, e.g., Refs.~\cite{W96,JR76,IZ80} and references therein.

The free Lagrangian $\mathcal{L}_0$ in the radiation gauge
gives rise to two modes of plane-wave solutions
with polarization vectors $\vec \epsilon_\pm (\vec k)$ and dispersion relations
\begin{eqnarray} \label{omegapm}
\omega_\pm(\vec k) =\left(  k^2 +\frac{1}{2}\,m^2 \pm \frac{1}{2}\, m\,
                           \sqrt{m^2 +4 k^2\cos^2 \theta }\,\right)^{1/2}\,,
\end{eqnarray}
with $k \equiv|\vec k|$ and $k\cos\theta \equiv \vec k\cdot \hat \eta $.
For generic wave vectors $\vec k$, the unit vectors
\begin{eqnarray} \label{xi123}
\hat \xi_1 \equiv (\hat \eta -\cos \theta\; \hat k)/\sin\theta \, , \quad
\hat \xi_2 \equiv (\hat k \times \hat \eta)/\sin\theta \,, \quad
\hat k     \equiv \vec k/|\vec k|\, ,
\end{eqnarray}
form an orthonormal tripod
with positive orientation, $\hat \xi_1 \times \hat \xi_2 =\hat k$.
The polarization vectors of the two modes are then given by
\begin{eqnarray}
\vec \epsilon_\pm (\vec k)= \frac{1}{\sqrt{2\Gamma (\Gamma \pm m)}}\,
\left[\, 2\,\omega_\pm \cos \theta \:\hat \xi_1 \mp i(\Gamma \pm m)\,\hat \xi_2\,
\right]\,,
\end{eqnarray}
with
\begin{eqnarray}
\Gamma = \Gamma(\vec k) \equiv \sqrt{m^2 +4k^2 \cos^2 \theta }\,.
\end{eqnarray}

As usual, the free quantized photon field can be defined in
terms of creation and annihilation operators,
\begin{eqnarray}
\vec A (x) &=&\int\frac{d^3 k}{(2\pi )^{3/2}}\;\left(\frac{1}{\sqrt{2\,\omega_+}}\,
   \left[\,  a_+  (\vec k)\, \vec \epsilon_+ (\vec k)\,
     \exp \left(-i\omega_+ t  +i\vec k \cdot \vec   x \right)
     +\mathrm{H.c.}\, \right] \right.\nonumber\\
&& \left. + \, \frac{1}{\sqrt{2\,\omega_-}}\,
      \left[\,  a_- (\vec k)\, \vec \epsilon_- (\vec k)\,
      \exp \left(-i\omega_- t +i\vec k \cdot\vec  x \right)
      +\mathrm{H.c.}\, \right]\right)\,.
\label{photonfield}
\end{eqnarray}
The creation and annihilation operators have standard commutation
relations,
\begin{eqnarray}
[\,a_s (\vec k) ,a_t (\vec l)\,]                =
[\,a^\dagger_s (\vec k) ,a^\dagger_t (\vec l)\,]= 0 \,, \;\;\;
[\,a_s (\vec k) ,a^\dagger_t (\vec l)\,]
=\delta_{s,t}  \: \delta^3 (\vec k -\vec l)\,,
\end{eqnarray}
for $ s,t \in \{+,-\}\,$.
The Fock space of photon states is then readily constructed
\cite{W96,JR76,IZ80}, starting from the vacuum state $|0\rangle$
with the property that $a_s (\vec k)\, |0\rangle  = 0$ for any $(\vec k, s)$.

The free electric and magnetic field operators can be expressed in the same
way as Eq.~(\ref{photonfield}), replacing the polarization
vectors $\vec \epsilon_\pm$ there by
\begin{eqnarray}
\vec f_\pm (\vec k) &\equiv& i \omega_\pm\, \vec \epsilon_\pm + m
  \left( \vec \epsilon_\pm \cdot (\hat k \times \hat \eta ) \right) \hat k\,=
   \frac{i}{\sqrt{ 2\Gamma (\Gamma \pm m)}}
\nonumber\\
&&  \times \left( 2\,\omega_\pm^{\, 2} \cos
     \theta \;\hat \xi_1 \mp i\omega_\pm (\Gamma \pm m) \,\hat \xi_2 \mp
     m\sin\theta \,(\Gamma \pm m)\,\hat k \right)
\label{f-pol}
\end{eqnarray}
for $\vec E$ and by
\begin{eqnarray} \label{b-pol}
\vec b_\pm (\vec k) \equiv i\vec k\times \vec \epsilon_\pm =
\frac{i k}{\sqrt{ 2 \Gamma (\Gamma \pm m)}} \left( 2\,\omega_\pm
\cos\theta\; \hat \xi_2 \pm i(\Gamma \pm m)\, \hat \xi_1 \right)
\end{eqnarray}
for $\vec B$.

Before continuing with the quantized theory, it may be useful
to consider the polarizations that result from Eqs.~(\ref{f-pol})
and (\ref{b-pol}) for the classical electric and magnetic fields
of a plane wave propagating \emph{in vacuo} (see also Ref.~\cite{K01}).

For $\theta =0$ (i.e., a wave vector $\hat k$ in the preferred direction $\hat\eta$),
the  $+$ and $-$ modes are circularly polarized.
But the vectors $\hat \xi_1$ and $\hat \xi_2$ from Eq.~(\ref{xi123}) are not
well-defined at $\theta =0$. Instead, they may be defined by a
limiting procedure, where the limit $k_\perp \downarrow \,0$
is performed for a wave vector
$\vec k =k_\perp\, \hat e_\perp + k_\parallel \, \hat \eta\,$.
Here, $\hat e_\perp$ is an arbitrary unit vector perpendicular to $\hat \eta $.
The resulting vectors are $\hat \xi_1 =-\hat e_\perp$ and
$\hat \xi_2 =\hat e_\perp \times \hat \eta$.
Consequently, the $+$ mode is right-handed
($R$, negative helicity) and the $-$ mode is
left-handed ($L$, positive helicity).

For $0<\theta <\pi/2$, both modes are elliptically polarized
and the helicities remain the same ($R$/$L$ for $+$/$-$). As long as
$k\cos \theta \gg m$, the modes are essentially circularly
polarized. In the opposite limit $k\cos \theta \ll m$, both modes approach
linear polarizations.

For $\theta =\pi/2$, the two modes behave somewhat
differently. The $-$ mode becomes exactly linearly polarized, with
$\vec f_- \sim  i\omega_- \hat \xi_1 \perp  \hat k$.
But the electric field for the $+$
mode has also a longitudinal component, $\vec f_+ \sim i\omega_+ \hat
\xi_2 +m \hat k$. This longitudinal component is, however,
relatively unimportant for frequencies $\omega_+ \gg m$.

For $\pi/2 <\theta <\pi$, both modes are again elliptically
polarized, but the helicities have changed ($L$/$R$ for $+$/$-$).
For $\theta =\pi$, finally, both modes are circularly polarized, with
helicities opposite to the $\theta =0$ case.

We now return to the quantized theory.
The energy-momentum ``tensor'' of the free theory is given by
\begin{eqnarray} \label{Thetamunu}
\Theta^{\mu\nu}= \frac{1}{4}\,g^{\mu\nu}F^{\rho\sigma}F_{\rho\sigma}
- F^{\mu\rho} F^{\nu\sigma} g_{\rho\sigma}
-\frac{m}{4}\:\eta^\nu \epsilon^{\mu\rho\sigma\tau}A_\rho F_{\sigma\tau} \,.
\end{eqnarray}
This tensor is conserved, $\partial_\mu \Theta^{\mu\nu}=0$, as may be checked
with the help of the equations of motion,
\begin{eqnarray}
\partial_\mu F^{\mu\nu}=
           \frac{m}{2}\; \epsilon^{\nu\rho\sigma\tau}\eta_\rho F_{\sigma\tau} \,.
\end{eqnarray}
The energy-momentum tensor (\ref{Thetamunu}) is, however,
not symmetric, as long as $m$ is nonzero.

For the case of a purely spatial \CS~``four-vector'' $\eta^\mu$,
the corresponding field energy and momentum are
\begin{eqnarray}
\mathcal{E}&\equiv& \int d^3 x \;\Theta^{00} =\int d^3x \;\frac{1}{2}
\left( \, |\vec E\,|^{\,2} + |\vec B\,|^{\,2} \,\right)\,,\\[1mm]
\mathcal{P}^i &\equiv& \int d^3 x \;\Theta^{0i} =\int d^3 x \;
 \left[\left(\vec E\times \vec B\,\right)^i  +\,\frac{m}{2}\:\hat\eta^i\,
       \left(\vec B \cdot \vec A\,\right)                          \right] \,.
\end{eqnarray}
These operators in terms of the quantized free photon field must, of course, be
normal-ordered \cite{W96,JR76,IZ80}. One finds after some algebra
\begin{eqnarray}
:\mathcal{E}: \; &=&\int d^3 k\;
                    \left[\, \omega_+(\vec k)\;a^\dagger_+ (\vec k) a_+ (\vec k) +
                    \omega_-(\vec k)\;a^\dagger_- (\vec k) a_- (\vec k)\, \right]\,,
\label{E} \\
:\vec{\mathcal{P}}: \; &=& \int d^3 k\; \,
\left[\,\vec k \; a^\dagger_+ (\vec k) a_+ (\vec k) +
        \vec k \;a^\dagger_- (\vec k) a_- (\vec k)\, \right]\,.
\label{vecP}
\end{eqnarray}
The energy eigenvalues, in particular, are nonnegative for the functions
$\omega_\pm$ as given by Eq.~(\ref{omegapm}).

Note, finally, that the Lagrangian (\ref{calL0}), with fixed parameters
$\eta^\mu =(0,\hat \eta)$, is translation invariant but not rotation invariant.
This implies that energy-momentum is conserved but not angular momentum.

\section{Kinematics of photon decay} \label{Kinematics}

In this section, we study the kinematics for the decay of one
initial photon with momentum $\vec q$ and energy $\omega_\pm (\vec q)$
into three final photons with energies
$\omega_\pm (\vec k_i)$,  $i=1, 2, 3$, and total momentum
$\vec k_1 +\vec k_2 +\vec k_3=\vec q$.  The theory considered
has been given in Section 2.
In a first reading, it is possible to skip ahead to the last paragraph of
this section, which summarizes the results.

For the usual Lorentz-invariant case ($m=0$),
the total energy of the final state is minimal when all
final momenta $\vec k_i$ are parallel to the initial momentum $\vec q$.
The reason is that adding perpendicular momenta,
$\vec k_i \to \vec k_i + \Delta \vec k_i$ with $\Delta \vec k_i \cdot \vec q =0$ and
$\sum_i \Delta \vec k_i =0$, increases the final energy, since
$\omega (\vec k_i +\Delta \vec k_i) >\omega (\vec k_i)$.

The same does not hold in our case ($m \neq 0$).
For infinitesimal $\Delta \vec k$ with $\Delta\vec k \cdot \vec k =0$, we find
\begin{eqnarray} \label{perp-mom}
\omega_\pm (\vec k +\Delta \vec k) \approx
\omega_\pm (\vec k) \pm \frac{m}{\omega_\pm\Gamma}\,
\left(\vec k \cdot \hat \eta \right) \left(\Delta \vec k \cdot \hat \eta \right),
\end{eqnarray}
which can be larger or smaller than $\omega_\pm (\vec k)$. This may
then result in the lowering of the total energy of a given 3-photon final state,
which can be shown as follows.

Assume that we start from a configuration where the final momenta
$\vec k_i$, $i=1,2,3$,  are parallel to the initial momentum $\vec q\,$,
which has $\vec q \cdot \hat\eta \neq 0$ and $\vec q \times \hat\eta \neq 0$.
Take the final particles 1 and 2 to have the same polarization ($+$ or
$-\,$) and leave the polarization of the final particle 3 unspecified ($\pm$).
Then add infinitesimal perpendicular momenta  $\Delta \vec k_i$
to the final momenta $\vec k_i$
and choose $\Delta \vec k_1 =-\Delta \vec k_2$ and $\Delta \vec k_3 = 0$,
so that overall momentum conservation is maintained.
According to Eq.~(\ref{perp-mom}), one of the first two final
energies is increased and the other decreased
(the third final energy is, of course, unchanged).
Moreover, the total final energy for the $(--\pm)$ case is decreased
if the energy corresponding to the smaller momentum is decreased
(for example, if $|\vec k_1|<|\vec k_2|$ then
$|\Delta \omega_- (\vec k_1)| > |\Delta \omega_- (\vec k_2)|\,$).
The same result is obtained for the $(++\pm)$ case,  as long as
$|\vec{k}_j \cdot \hat\eta| > m/\sqrt{2}$ for $j=1,2$.

Still, it is worthwhile to study the case of parallel (collinear)
final momenta in some detail. The frequencies $\omega_\pm$ may
then be treated as functions of \emph{scalar} variables,
$\omega_\pm=\omega_\pm (k_i)$ for $k_i \equiv |\vec k_i|$.
Assuming both $k_1$ and $k_2$ to be nonzero, one finds  for the
$-$ state that
\begin{eqnarray} \label{ineq-}
\omega_- (k_1 +k_2) > \omega_-(k_1) +\omega_- (k_2)\,,
\end{eqnarray}
except for the special case of $\vec k_1 \cdot\hat\eta =
\vec k_2 \cdot\hat\eta =0$, for which Eq.~(\ref{ineq-}) becomes an equality.
As a consequence, the decay $- \to ---$ is allowed kinematically. This also
implies that the decay $+\to ---$ is allowed,
because $\omega_+ (q) >\omega_- (q)$.

For the $+$ state, on the other hand, the following inequality  holds:
\begin{eqnarray} \label{ineq+}
\omega_+ (k_1 +k_2) <\omega_+
(k_1) +\omega_+ (k_2)\,,
\end{eqnarray}
again assuming nonzero $k_1$ and $k_2$. In this case,
no immediate conclusions can be drawn from the inequality,
because of relation (\ref{perp-mom}).
The inequalities (\ref{ineq-}) and (\ref{ineq+}) are proven in
Appendix  \ref{Energyinequalities}.

Another kinematically allowed decay is $+\to +--$. This can be shown by
establishing the following inequality for appropriate parallel momenta:
\begin{eqnarray} \label{ineq+--}
\omega_+ (k_1 +k_2 +k_3) > \omega_+ (k_1) + \omega_- (k_2) +\omega_- (k_3) \,,
\end{eqnarray}
with the notation $k_i \equiv |\vec k_i|$. Unlike the
inequalities (\ref{ineq-}) and (\ref{ineq+}), inequality (\ref{ineq+--})
does not hold for arbitrary values of the $k_i$.
Instead, some restrictions must be imposed on the momenta.

Concretely, assume that $k_2 \ll  m$, $k_2 \ll k_1$ and $k_3 \ll  m$, $k_3
\ll k_1$. Expanding the \lhs~of Eq.~(\ref{ineq+--}) up to first order in $k_2
+k_3$ and the \rhs~up to first order in $k_2$ and $k_3$, we find
\bdi
\omega_+^{-1} (k_1)\left( k_1 +\frac{2mk_1 \cos^2 \theta}{\sqrt{m^2 +4k_1^2
\cos^2 \theta}} \,\right) (k_2 +k_3) > \sin\theta \, (k_2 +k_3)\,.
\edi
Defining $x\equiv m/k_1$, this inequality can be written as
\bdi 1+\frac{2\,x\cos^2 \theta}{\sqrt{x^2 +4\cos^2 \theta}} >
\sin \theta \left( 1+\frac{x^2}{2} +\frac{x}{2}\,\sqrt{x^2+4\cos^2 \theta}
           \,\right)^{1/2} .
\edi
Both sides of the last inequality are manifestly
positive. Squaring both sides and re-arranging them somewhat, we arrive at
\bdi -x^2 +2\cos^2 \theta +\frac{8\,x^2 \cos^4 \theta}{x^2 +4\cos^2 \theta} >
x\:\frac{x^2 -4\cos^2\theta}{\sqrt{x^2 +4\cos^2\theta}} \,.
\edi
This final inequality certainly holds if the \lhs~is larger than zero {\em and}
the \rhs~smaller than zero. Writing $b\equiv 4\cos^2 \theta$, the
condition for the \lhs~gives
\bdi x^2 < b\,(b-1)/4+  (b/4)\,\sqrt{b^2 -2\,b+9}
\edi
and the condition for the \rhs
\bdi x^2 <b\,.
\edi
Both conditions can be fulfilled simultaneously, as long as $\cos^2 \theta \neq 0$.
We might, for example, restrict the momenta, so that $b>2$, which corresponds to
$\cos^2 \theta > 1/2$. Then both conditions are equivalent and give
$x<\sqrt{2}$, which corresponds to $k_1 >m/\sqrt{2}$.
[For the special case of $\cos^2 \theta =0$,
the quantity $\omega_+ $ ($\omega_-$) is effectively the energy of a massive
(massless) particle, which implies that the
inequality (\ref{ineq+--}) does not hold for small enough values of $k_i/m$.]

The above discussion shows that the energy inequality (\ref{ineq+--}) holds
for certain restricted values of the final collinear momenta.
By continuity, this energy inequality holds also if the final momenta
deviate slightly from collinearity. Therefore, our restrictions are
restrictions to certain \emph{regions} in phase space and not
restrictions to submanifolds of measure zero. As a consequence,
the decay $+\to +--$ is allowed kinematically.

Altogether, we have shown that the three decay channels $-\to ---$,
$+\to ---$ and $+\to +--$ are allowed kinematically. For the five
other decay channels, we have not been able to find allowed regions in
phase space, either analytically or numerically. These allowed
regions in phase space perhaps do not exist, but this remains to
be proven.

\section{Matrix element $+\to - - -$}

We now calculate the matrix element for the decay of a $+$ polarization mode
into three $-$ modes.
To lowest order, there are two contributions to this matrix element
from the quartic Euler--Heisenberg Lagrangian.
The  first term in Eq.~(\ref{calL1}) gives the following contribution:
\begin{eqnarray}
&&
\langle 0 |\, a_+ (\vec q)\, :\!\int d^4 x\;
\left( \, |\vec E (x)|^{\,2} -  |\vec B (x)|^{\,2} \,\right)^2\!:
\,a^\dagger_- (\vec k_1) a^\dagger_- (\vec k_2) a^\dagger_- (\vec k_3)\,
|0\rangle =
\no \\&&
- \,\frac{1}{2\, \pi^2} \; \delta [\,\omega_+ (\vec q) -\omega_- (\vec k_1) -
 \omega_- (\vec k_2) -  \omega_- (\vec k_3)\,] \, \,
\delta^3 (\vec q -\vec k_1 - \vec k_2 -\vec k_3 )
\no \\&&
\times \left[\, \omega_+ (\vec q) \, \omega_- (\vec k_1) \, \omega_- (\vec k_2)
\, \omega_- (\vec k_3) \, \right]^{-1/2}
\no \\&&
\times
\Bigl(
\left[\vec f_- (\vec k_1) \cdot \vec f_- (\vec k_2)
       - \vec b_- (\vec k_1) \cdot \vec b_- (\vec k_2) \right]\,
\left[  \vec f_- (\vec k_3) \cdot \vec f_+^* (\vec q)
       -\vec b_- (\vec k_3) \cdot \vec b_+^* (\vec q)      \right]
\no \\&&
+\phantom{\Bigl(}\!
\left[    \vec f_- (\vec k_1) \cdot \vec f_- (\vec k_3)
        - \vec b_- (\vec k_1) \cdot \vec b_- (\vec k_3)\right]\,
\left[  \vec f_- (\vec k_2) \cdot \vec f_+^* (\vec q)
      - \vec b_- (\vec k_2) \cdot \vec b_+^* (\vec q)             \right]
\no \\&&
+\phantom{\Bigl(}\!
\left[\vec f_- (\vec k_2) \cdot \vec f_- (\vec k_3)
       - \vec b_- (\vec k_2) \cdot \vec b_- (\vec k_3) \right]\,
\left[  \vec f_- (\vec k_1) \cdot \vec f_+^* (\vec q)
      - \vec b_- (\vec k_1) \cdot \vec b_+^* (\vec q)       \right]
\Bigr) \,.\no \\&&
\label{matr1}
\end{eqnarray}
The second term in Eq.~(\ref{calL1}) gives, with different combinatorics,
\begin{eqnarray} &&
\langle 0 |\, a_+ (\vec q)\,
:\!\int d^4 x\; 7\left(\vec E (x)\cdot\vec B (x)\right)^2\!:
\, a^\dagger_- (\vec k_1) a^\dagger_- (\vec k_2) a^\dagger_- (\vec k_3)\,
|0\rangle =
 \no \\&&
-\,\frac{1}{2\, \pi^2} \; \delta [\,\omega_+ (\vec q) -\omega_- (\vec k_1) -
 \omega_- (\vec k_2) -  \omega_- (\vec k_3)\,] \, \,
\delta^3 (\vec q -\vec k_1 - \vec k_2 -\vec k_3 )
 \no \\&&
\times \left[\,  \omega_+ (\vec q) \, \omega_- (\vec k_1) \, \omega_- (\vec k_2)
\, \omega_- (\vec k_3)\, \right]^{-1/2} \times 7/4
 \no          \\&&
\times \Bigl(
\left[  \vec f_- (\vec k_1) \cdot \vec b_- (\vec k_2)
      + \vec f_- (\vec k_2) \cdot \vec b_- (\vec k_1) \right]\,
\left[  \vec f_+^* (\vec q) \cdot \vec b_-   (\vec k_3)
      + \vec f_- (\vec k_3) \cdot \vec b_+^* (\vec q)  \right]
 \no   \\&&
+\phantom{\Bigl(}\!
\left[   \vec f_- (\vec k_1) \cdot \vec b_- (\vec k_3)
       + \vec f_- (\vec k_3) \cdot \vec b_- (\vec k_1)\right]\,
\left[ \vec f_+^* (\vec q) \cdot \vec b_- (\vec k_2)
      +\vec f_- (\vec k_2) \cdot \vec b_+^* (\vec q)  \right]
 \no \\&&
+\phantom{\Bigl(}\!
\left[   \vec f_- (\vec k_2) \cdot \vec b_- (\vec k_3)
       + \vec f_- (\vec k_3) \cdot \vec b_- (\vec k_2)\right]\,
\left[ \vec f_+^* (\vec q) \cdot \vec b_- (\vec k_1)
      +\vec f_- (\vec k_1) \cdot \vec b_+^* (\vec q)  \right]
\Bigr) \,.\no \\&&
\label{matr2}
\end{eqnarray}

Unfortunately, the expressions involving the polarization vectors in
Eqs. (\ref{matr1})--(\ref{matr2}) are rather complicated
and  one would like to simplify them by using some type of small $m$ expansion.
But such a procedure appears to be quite difficult and, in this paper,
we keep the full expressions.

The probability for decay of a single $+$ state into three $-$ states can
be calculated by
integrating the square of the amplitude over the final momenta.
The result will, however, be a \emph{function} of
$|\vec q|$ and $\hat q\cdot\hat \eta$, for given $m$ and $M$.
For this reason, we turn  to a simpler problem in the next section, namely
the decay of a $+$ state at rest.

\section{Partial decay width for $+\to - - -$}

The decay probability for $+\to - - -$ will be evaluated for the case of
vanishing initial momentum, $\vec q=0$.
The energy $\omega_+ (\vec q)$ for $\vec q =0$ equals $m$ and the
initial ``photon'' behaves like a massive particle at rest.
The partial decay width is then obtained by
squaring the matrix element of the previous section and
integrating over the final momenta $\vec k_i$, for $i=1\ldots 3$.
A combinatorial factor $1/3!=1/6$
must be inserted because of the three identical particles in the final state.

Purely on dimensional grounds, we can write the partial decay width to order
$\alpha^4$ as
\begin{eqnarray} \label{Gamma+to---}
\Gamma_{+\to - - -}^{\,(4)} =
\frac{1}{512\,\pi^5}\; \kappa \;
\left(\frac{2\, \alpha^2}{45}\right)^2 \;\frac{m^9}{M^8}\,,
\end{eqnarray}
with a \emph{single number} $\kappa \geq 0$ to be determined.

The relevant phase space integral is defined as follows:
\mathindent=1em
\begin{eqnarray}  \label{int---}
I_{---} &\equiv& \int\, d^3 k_1\, d^3 k_2\, d^3 k_3\;
\delta^3 (\vec k_1  + \vec k_2 +\vec k_3 ) \,
\delta [\,\omega_- (\vec k_1) + \omega_- (\vec k_2) + \omega_- (\vec k_3)-m\,]
 \no\\
&&
\times \left[\, \omega_+ (0) \, \omega_- (\vec k_1) \, \omega_- (\vec k_2)
 \, \omega_- (\vec k_3)\,\right]^{-1}\;
 g(\vec k_1,\vec k_2,\vec k_3) \,,
\end{eqnarray}
\mathindent=2em  
where the  nonnegative function $g$ depends on the
electric and magnetic polarization vectors.
Specifically, the factor $g$ is given by the
absolute value squared of the terms in large brackets in
Eqs.~(\ref{matr1}) and (\ref{matr2}) for $\vec q=0$,
\mathindent=0em
\begin{eqnarray} \label{fdef}
g(\vec k_1,\vec k_2,\vec k_3) &=& \left|
\Bigl(
\left[\vec f_- (\vec k_1) \cdot \vec f_- (\vec k_2)
- \vec b_- (\vec k_1) \cdot \vec b_- (\vec k_2) \right]\,
\vec f_- (\vec k_3) \cdot \vec f_+^* (0) + \cdots \Bigr) +
\frac{7}{4} \right. \no\\
&& \left. \times \Bigl(
\left[\vec f_- (\vec k_1) \cdot \vec b_- (\vec k_2)
+ \vec f_- (\vec k_2) \cdot \vec b_- (\vec k_1) \right]\,
  \vec f_+^* (0)      \cdot \vec b_- (\vec k_3) + \cdots \Bigr) \right|^{\, 2} \,. \no\\
&&
\end{eqnarray}
\mathindent=2em
The expressions for $\vec f_+ (0)$ and $\vec b_+ (0)$ will be given in the
next section. Note that only the mass-scale $m$ appears in the integral (\ref{int---}),
so that $I_{---} \propto m^9$.

A numerical calculation shows that the integrand of Eq.~(\ref{int---})
is strictly positive over a finite region of phase space (see
Appendix \ref{Numericalresults}).
For the numerical constant $\kappa $ in the partial decay width
(\ref{Gamma+to---}), this implies
\begin{eqnarray}  \label{kappabound}
\kappa =  \left( \frac{1}{3\,!} \right) \, \left( \frac{1}{2\, \pi^2} \right)^2 \;
        I_{---}\,/\, m^{9} \,> 0 \,,
\end{eqnarray}
which is the main result of the present paper.

\section{Partial decay width for $+\to + - -$}

The decay of an initial $+$ polarization state is in general more
complicated than the decay of an initial $-$ state.
But for initial momentum $\vec q=0$ a major
simplification occurs: the second decay channel $+\to +-- $ no
longer contributes, as will be demonstrated in this section.

The matrix element for $+\to + - -$ decay follows from
Eqs.~(\ref{matr1}) and (\ref{matr2})
by replacing \emph{all} $-\,$ labels that occur in conjunction with $\vec k_1\,$
by $+$ labels. The energy delta function, for example, becomes
\begin{eqnarray}
 \delta [\,\omega_+ (\vec q) -\omega_+ (\vec k_1) -
 \omega_- (\vec k_2) -  \omega_- (\vec k_3)\,]\,.
\end{eqnarray}

The energy conservation condition for the decay of a $+$ photon at
rest ($\vec q =0$) now takes the form
\begin{eqnarray}
\left(\omega_+ (\vec k_1)-m\right) + \omega_- (\vec k_2) + \omega_- (\vec k_3) =0\,.
\end{eqnarray}
This relation holds only if all three final momenta $\vec k_i$ are zero,
because each of the three terms in the above equation is positive
semi-definite and zero only for vanishing momentum.
The conditions $\vec k_i =0$,
for $i=1,2,3$, are restrictions to a lower-dimensional submanifold in phase space
and have, therefore, measure zero. The resulting phase space integrals will
lead to a zero contribution to the decay width, unless certain singularities
of the phase-space integrand appear in the limit $\vec k_i \to 0$.
In the following, we will show that such infrared singularities are absent.

The phase-space integral in spherical polar
coordinates $k_i,\theta_i,\varphi_i$,  for $i=1,2,3$, takes the form
\begin{eqnarray}  \label{int+--}
I_{+--} &\equiv& \int \delta^3 (\vec k_1  + \vec k_2 +\vec k_3 )
\, \delta [\, \omega_+ (\vec k_1) + \omega_- (\vec k_2) + \omega_-
(\vec k_3)-m\,] \; h(k_i ,\theta_i ,\varphi_i) \no\\
&& \times\;\frac{ k_2^2 \, k_3^2 \, \sin\theta_2\, \sin\theta_3 \;
 d^3 k_1 \, dk_2 \, dk_3 \,  d \theta_2 \, d \theta_3 \, d\varphi_2 \,
 d \varphi_3 }{\omega_+ (0) \, \omega_+ (\vec k_1) \, \omega_- (\vec k_2)
 \, \omega_- (\vec k_3)}\,,
\end{eqnarray}
where the  nonnegative
function $h(k_i ,\theta_i ,\varphi_i)$ depends on the electric
and magnetic polarization vectors $\vec f_\pm (\vec k_i)$ and
$\vec b_\pm (\vec k_i)$.
Using the bounds $\omega_+ (\vec k) \ge m$ and $\omega_- (\vec k)\ge k\sin\theta$
in the denominator and integrating over the momentum $\vec k_1$,
we find for the phase-space integral
\begin{eqnarray} \label{int+--k2k3}
0\le I_{+--} \le  &&\int \delta [\, \omega_+ (\vec k_2 +\vec k_3)
+ \omega_- (\vec k_2) + \omega_- (\vec k_3)-m\,] \,  h(k_j
,\theta_j ,\varphi_j) \no\\
&& \times\; \frac{ k_2 \, k_3 \; dk_2 \, dk_3 \,
             d \theta_2 \, d \theta_3 \, d\varphi_2 \,d \varphi_3 }{m^2}\,,
 \label{int+--2}
\end{eqnarray}
where $h$ now depends on the polarization vectors $\vec f_\pm (\vec k_j)$ and
$\vec b_\pm (\vec k_j)$, for $j=2,3$, and the vectors
$\vec f_\pm (-\vec k_2  -\vec k_3)$ and $\vec b_\pm (-\vec k_2  -\vec k_3)$.

The integral in Eq.~(\ref{int+--k2k3}) has
two potential sources of infrared singularities,
the first of which is the factor   $h(k_j ,\theta_j ,\varphi_j)$.
But it is not difficult to see that $h(k_j ,\theta_j ,\varphi_j)$ is
nonsingular in the limit $k_j \to 0$. As $h$ depends only
on the electric and magnetic polarization vectors, it suffices to
demonstrate that these vectors are nonsingular in the infrared.
Writing $\vec k = k_\perp \,\hat e_\perp + k_\parallel \,\hat \eta$
with $\hat e_\perp$ an arbitrary unit vector orthogonal to $\hat \eta$,
it can indeed be shown that
\begin{eqnarray}
\lim_{k \to 0}\, \vec f_+ (\vec k) = i m (\hat e_\perp -i \hat
e_\perp \times \hat \eta )
\end{eqnarray}
and
\begin{eqnarray}
\lim_{k \to 0}\, \vec f_- (\vec k) = \lim_{k \to 0}\, \vec b_+ (\vec k) =
\lim_{k \to 0}\, \vec b_- (\vec k) =0\,.
\end{eqnarray}

The second potential  source of infrared singularities in the integral
(\ref{int+--k2k3}) is the energy delta function itself,
if singularities of the type $\int dk\, \delta (k^2)$
are produced. The proof that this does not happen is
somewhat involved and is relegated to Appendix  \ref{Analyticresult}.

Altogether, we find that the $+\to +--$ channel does not contribute
to $+$ decay at vanishing initial momentum, $\vec q =0$, so that
\begin{eqnarray} \label{Gamma+to+--}
\Gamma_{+\to +--}^{\,(4)} =0\,.
\end{eqnarray}
For $\vec q=0$, no other channels contribute, besides the two channels already
considered. The reason is energy conservation (cf. Section 2):
the energy of an initial $+$ state at rest is $m$, whereas the energies
of the final states $++-$ and $+++$ are at least $2\,m$ and $3\,m$,
respectively.

\section{Total decay rate of a $+$ photon at rest}

In this paper, we have studied the decay of photons in a relatively simple
model, for which the Lagrangian (\ref{calL})
contains only the usual Maxwell and Euler--Heisenberg terms, together
with  a  CPT-violating \CS-like term.
The two mass parameters of the model are $m$, which enters the \CS-like term
(\ref{calL0}) linearly, and $M$, whose fourth power enters the Euler--Heisenberg
term (\ref{calL1}) inversely. The Euler--Heisenberg term has, in addition, an
overall coupling constant $\alpha^2$.

The photon of this model has two polarization states, labeled $\pm$, with
energies and momenta given by Eqs.~(\ref{omegapm}), (\ref{E}) and
(\ref{vecP}). The total decay rate to order $\alpha^4$
of a $+$ state at rest follows from the sum of the
partial decay widths (\ref{Gamma+to---}) and (\ref{Gamma+to+--}),
\begin{eqnarray} \label{Gamma+}
\Gamma_{+}^{\,(4)} =
\frac{1}{512\,\pi^5}\; \kappa \;\left(\frac{2\, \alpha^2}{45}\right)^2 \;
\frac{m^9}{M^8}\,,
\end{eqnarray}
with a constant $\kappa \geq 0$.
A numerical calculation shows that $\kappa$ is nonvanishing and, most likely, not
vastly different from $1$.
[See Eq.~(\ref{kappabound}) and Appendix \ref{Numericalresults}.]

As explained in Section 2, the model considered is part of the effective
gauge field action of quantum electrodynamics with an additional \CS-like
term. An order of magnitude estimate for the photon mean lifetime can then
be obtained by writing
\begin{eqnarray} \label{Gamma+inverseyrs}
\Gamma_{+}^{\,(4)} = \kappa\, \left[\,3\;10^{336}\;\mathrm{yr} \,\right]^{-1}\,
\left( \frac{\alpha}{1/137} \right)^4\,
\left( \frac{511\;\mathrm{keV}}{M} \right)^8\,
\left( \frac{m}{10^{-33}\;\mathrm{eV}} \right)^9\,,
\end{eqnarray}
where for $\alpha$ and $M$ the values of the fine-structure constant and the
electron mass have been inserted and for $m$ the experimental
upper limit from polarization measurements on distant radio galaxies
(see Refs.~\cite{CFJ90,WPC97} and references therein).
For $m \sim 10^{-35}\;\mathrm{eV}$ as might be expected from the CPT anomaly
\cite{K00}, the photon lifetime would be larger by a factor of $10^{18}$.

The very large photon lifetime as indicated by Eq.~(\ref{Gamma+inverseyrs})
is perhaps not of direct relevance, at least for the current epoch in the history
of the universe.
There remain, however, fundamental questions about the formulation of
this particular CPT-violating extension of quantum electrodynamics if
certain photon states are no longer absolutely stable (cf. Ref.~\cite{V63}).

\vspace{-0.25\baselineskip}
\section*{Acknowledgements}
\vspace{-0.25\baselineskip}
F.R.K. thanks M. Frank and J. Schimmel for help with the computer.
C.A. acknowledges support from the Austrian START award project
FWF-Y-137-TEC of N.J. Mauser.

\appendix
\vspace{-0.25\baselineskip}
\section{Energy inequalities} \label{Energyinequalities}
\vspace{-0.25\baselineskip}

In this appendix, we prove the energy inequalities (\ref{ineq-}) and (\ref{ineq+})
for parallel momenta $\vec k_1\, || \,\vec k_2$. [The $\omega_- $
case has the additional conditions $\vec k_j \cdot \hat \eta \neq 0$, for $j=1,2$.]
It is, in fact, not difficult to verify
that these inequalities hold for sufficiently small $k_j \equiv
|\vec k_j| \ll m$. For the general case, we give a proof by contradiction.

Assume that the inequalities do
\emph{not} hold for all values of $k_1$ and $k_2$. Then there must exist
values for $k_1$ and $k_2$ with
\begin{eqnarray} \label{omegaequal}
\omega (k_1 +k_2)=\omega (k_1) + \omega (k_2)\,.
\end{eqnarray}
[These special values of $k_1$ and $k_2$ could, of course,
be different for the $+$ and the $-$ case. But the proof is analogous in
both cases and we treat both cases at once, writing $\omega$ for
either $\omega_+$ or $\omega_-$.] The equality (\ref{omegaequal}) can also
be written as
\begin{eqnarray} \label{eq+-}
v_\mathrm{ph}(k_1 +k_2)=\frac{k_1}{k_1 +k_2}\,v_\mathrm{ph} (k_1) +
                        \frac{k_2}{k_1 +k_2}\,v_\mathrm{ph} (k_2)\,,
\end{eqnarray}
where $v_\mathrm{ph} (k)
\equiv \omega (k)/k$ is the absolute value of the phase velocity.

Equation (\ref{eq+-}), now, holds only if one of the
following three conditions is met:
\begin{eqnarray}
\mathrm{(i)}  &\;\;\;& v_\mathrm{ph} (k_1) =v_\mathrm{ph} (k_2) =v_\mathrm{ph} (k_1 +k_2)\,,
\no \\
\mathrm{(ii)} && v_\mathrm{ph} (k_1) > v_\mathrm{ph} (k_1 + k_2) \quad \mbox{and}\quad
v_\mathrm{ph} (k_2) < v_\mathrm{ph} (k_1 + k_2)\,,
\no \\
\mathrm{(iii)}&& v_\mathrm{ph} (k_1) < v_\mathrm{ph} (k_1 + k_2) \quad \mbox{and}\quad
v_\mathrm{ph} (k_2) > v_\mathrm{ph} (k_1 + k_2)\,.
\end{eqnarray}
Each of these conditions implies that $v_\mathrm{ph}(k) $ has an extremum
somewhere in the momentum interval $[\mbox{min}(k_1, k_2),k_1 +k_2]$.
But this conclusion contradicts the simple observation that,
for both $\omega_+$ and $\omega_-$,
the derivative of the phase velocity $dv_\mathrm{ph}/dk$
is nonzero for all positive values of $k$.
[The observation requires for the $\omega_-$ case the additional condition
$\vec k \cdot \hat \eta \neq 0$.]
This then implies that the assumption above Eq.~(\ref{omegaequal})
must be incorrect and that
the inequalities (\ref{ineq-}) and (\ref{ineq+}) hold for \emph{all}
positive values of $k_1$ and $k_2$, as stated in the main text.

\vspace{-0.25\baselineskip}
\section{Numerical result for $I_{---}$} \label{Numericalresults}
\vspace{-0.25\baselineskip}

In this appendix, we report on a numerical calculation of the
integral $I_{---}$, as defined by Eq.~(\ref{int---}).
A dimensionless quantity $I$ is obtained by setting $I \equiv I_{---}/m^9$.
We can be relatively
brief in describing our results since the calculation of a decay
rate is well-known (see, e.g.,  Sections~3.6 and 3.7 of Ref.~\cite{WS86}).
We proceed in four steps.

First, the integral over $\vec k_3$ is performed and the remaining six
integration variables are taken to be the following dimensionless
spherical coordinates:
\begin{eqnarray} \label{6coordinates}
k_1\equiv |\vec k_1|\, ,\, k_2\equiv |\vec k_2|\, ,\, \theta_1\, ,\, \theta_2\, ,\,
\varphi_\pm \equiv \varphi_2 \pm \varphi_1 \,,
\end{eqnarray}
with polar angles $\theta_1, \theta_2 \in [0,\pi]$
defined with respect to an axis in the preferred direction $\hat\eta$ and
azimuthal angles $\varphi_1,  \varphi_2 \in [0,2\pi]$  in the plane orthogonal to
this axis. The angles $\varphi_+$ and  $\varphi_-$ can be taken to run
over $[0,4\pi]$ and $[-\pi,+\pi]$, respectively.

Second, the energy delta function effectively sets $\varphi_-$ to a fixed value,
\begin{eqnarray} \label{varphimin}
 \varphi_- = (1-2\,n)\,\arccos\chi \,,
\end{eqnarray}
for a known function $\chi=\chi(k_1,\theta_1,k_2,\theta_2)$ and
integer $n = 0$ or $1$.
Also, there are the constraints that both $|\chi|$ and
$\omega_-(k_1,\theta_1)+\omega_-(k_2,\theta_2)$
must be less than 1, which can be implemented by
introducing appropriate step functions $\Theta(1-|\chi|)$ and
$\Theta[\, 1-\omega_-(k_1,\theta_1)-\omega_-(k_2,\theta_2)\, ]$
into the integrand.

Third, the resulting integral has the following structure:
\mathindent=0.5em
\begin{eqnarray} \label{I}
I = \int_{0}^{\infty}
    dk_1 \, dk_2  \int_{0}^{\pi} d \theta_1 \, d \theta_2
    \int_{0}^{4\pi}\,  \frac{d\varphi_+}{2} \, \sum_{n=0}^{1}\,
    \left. \mathcal{I} \:\right|_{\,\varphi_- = (1-2\,n)\,\arccos\chi\,;\,
                      \vec k_3 = -\vec k_1 -\vec k_2} \,,
\end{eqnarray}
\mathindent=2em
with the integrand
\begin{eqnarray} \label{int}
\mathcal{I} &=& \left[\,k_1/\omega_-(k_1,\theta_1)\,\right]\,
                \left[\,k_2/\omega_-(k_2,\theta_2)\,\right]\,
                \Theta[\, 1-\omega_-(k_1,\theta_1)-\omega_-(k_2,\theta_2)\, ]
\no\\[1mm]
&&\times\,\Theta\left(1-\left|\,\chi(k_1,\theta_1,k_2,\theta_2)\right|\,\right)\,
        \left|\,\sin\varphi_- \right|^{-1}\, g(\vec k_1,\vec k_2,\vec k_3)\,,
\end{eqnarray}
where $\omega_-$ and $g$ are dimensionless functions (i.e., the
expressions of the main text with $m\equiv 1$).
The integral (\ref{I}) is complicated, but its integrand is still
nonnegative; cf. the definition (\ref{fdef}).

Fourth, a numerical calculation with \textsc{Mathematica}  \cite{W99}
shows the integrand of Eq.~(\ref{I}) to be independent of $\varphi_+$ and $n$.
This effectively reduces the integral to a four-dimensional one and a
numerical estimate gives
\begin{eqnarray} \label{Iestimate}
I \equiv I_{---}/m^9 \approx \, 0.2 \,.
 \end{eqnarray}
From this estimate, one obtains the result (\ref{kappabound}) quoted in the main
text.

The value (\ref{Iestimate}) is to be considered preliminary.
More work is needed to obtain an accurate result, both analytically
(e.g., to make the independence of the azimuthal coordinate $\varphi_+$
manifest) and numerically (e.g., to sample phase space efficiently).

\vspace{-0.25\baselineskip}
\section{Analytic result for $I_{+--}$} \label{Analyticresult}
\vspace{-0.25\baselineskip}

In this appendix, we demonstrate the vanishing of the integral $I_{+--}$
as defined by Eq.~(\ref{int+--}). We start with two preliminary steps.
First, we introduce the following representation
for the energy delta function:
\begin{eqnarray}
\delta (\omega) =\lim_{a \to 0}\; \frac{1}{ \sqrt{\pi}\,a}\,
                  \exp \left(-\omega^2 /a^2\right)\, , \quad a>0 \, .
\end{eqnarray}

Second, we replace the nonnegative function $h(k_j,\theta_j ,\varphi_j )$
in the integral of Eq.~(\ref{int+--2}) by the following bound:
\begin{eqnarray} \label{h-bound}
h(k_j,\theta_j ,\varphi_j ) \le \sum_{l,n=0}^4 H(l,n)\, k_2^l\, k_3^n \,,
\end{eqnarray}
where $H(l,n)$ are appropriate nonnegative numbers.
This bound may be understood from the observation that the factor
$h(k_j,\theta_j ,\varphi_j )$ is the absolute square of a sum of terms of the
type
\begin{eqnarray} \label{vvvv}
[\vec v_1 (\vec k_2) \cdot \vec v_2 (\vec k_3)]\,
[\vec v_3 (-\vec k_2 -\vec k_3)\cdot \vec v_4 (0)]\,,
\end{eqnarray}
where  each vector $\vec v_i$, for $i=1,\ldots,4$, is either an electric
polarization vector $\vec f_\pm$ or  a magnetic polarization vector
$\vec b_\pm$, as defined by Eqs.~(\ref{f-pol}) and (\ref{b-pol}).
Using the energy bounds $\omega_+ (\vec k_j)\le k_j+m$ and
$\omega_- (\vec k_j) \le k_j$,
each vector $\vec v_i (\vec k_j)$
can be shown to obey the following inequality:
\begin{eqnarray}\label{vbound}
\left|\, \vec v_i (\vec k_j)\right| \le
\left|\, \vec v_{0,i}(\vec k_j) + k_j \,\vec v_{1,i}(\vec k_j)\right|\,,
\end{eqnarray}
where $\vec v_{0,i}$ and $\vec v_{1,i}$ are
uniformly bounded vectors (i.e., they never exceed a certain length).
In the product (\ref{vvvv}), then, the momenta $k_2$ and $k_3$ show up
at most quadratically, so that no powers higher than four occur
in the absolute square of a sum of such terms.
This explains the bound (\ref{h-bound}).

With these two steps, the bound (\ref{int+--2}) becomes
\mathindent=0em
\begin{eqnarray} \label{initialbound}
0\le I_{+--} \le &&\lim_{a \to 0}\; \frac{1}{a\sqrt{\pi}\,m^2 }
\int_{0}^{\infty} dk_2 \, dk_3 \, \int_{0}^{\pi} d \theta_2 \, d \theta_3 \,
\int_{0}^{2\pi} d\varphi_2 \,d \varphi_3\;
\sum_{l,n=0}^4 H(l,n)
\no\\
&& \times \, k_2^{l+1}\, k_3^{n+1}\,
   \exp \Bigl( -\left[\,\omega_+ (\vec k_2 +\vec k_3)-m + \omega_-
   (\vec k_2) +
                     \omega_- (\vec k_3)\right]^2 /a^2 \Bigr)\,.\no\\
&&
\end{eqnarray}
\mathindent=2em
Now change the momentum variables $k_j$ to $k_j/a$ and use the estimates
$\omega_+  \ge m$ and $[\omega_- (k_2) +\omega_- (k_3)]^2 \ge \omega_-^2
(k_2)+ \omega_-^2 (k_3)$. This gives
\begin{eqnarray}
0\le I_{+--} \le &&\lim_{a \to 0}\;  \sum_{l,n=0}^4
\frac{a^{3+l+n}}{\sqrt{\pi}\,m^2 } \int
 dk_2 \, dk_3 \,  d \theta_2 \, d \theta_3 \, d\varphi_2 \,d \varphi_3\;
 H(l,n)\,k_2^{l+1}\, k_3^{n+1}
\no\\
&&
\times \exp \Bigl(
-k_2^2 - \frac{m_a^2}{2}
+ \frac{m_a^2}{2}\, \sqrt{1+4\cos^2\theta_2\, k_2^2/m_a^2}
\no\\
&&
-k_3^2 - \frac{m_a^2}{2}
+ \frac{m_a^2}{2}\, \sqrt{1+4\cos^2 \theta_3 \,k_3^2/ m_a^2}\: \Bigr)\,,
\end{eqnarray}
with $m_a \equiv m/a$. Since $\cos^2 \theta_j \le 1$, we finally arrive at
\begin{eqnarray} \label{finalbound}
0\le I_{+--} \le &&
\lim_{m_a \to \infty}\;  \sum_{l,n=0}^4 H(l,n)\,
\frac{m^{1+l+n}}{\sqrt{\pi}\,m_a^{3+l+n} }
\int_{0}^{\pi} d \theta_2 \, d \theta_3 \,\int_{0}^{2\pi} d\varphi_2 \, d \varphi_3 \;
I_l \, I_n \, , \nonumber \\
&&
\end{eqnarray}
in terms of the dimensionless integrals
\begin{eqnarray} \label{Il}
I_l &\equiv& \int_0^\infty d k \; k^{l+1} \exp \Bigl(
-k^2  -\frac{m_a^2}{2} + \frac{m_a^2}{2}\, \sqrt{1+4 k^2/m_a^2}\:\Bigr)\,.
\end{eqnarray}

The momentum integrals $I_l$, for $l=0,\ldots,4$,
can be evaluated analytically with the help of \textsc{Mathematica}
\cite{W99}, but we are only interested in their asymptotic behavior for
$ m_a\to\infty$. Making the change of variables $y=k^2/m_a$ and Taylor
expanding the square root in Eq.~(\ref{Il}), we obtain
\begin{eqnarray}
 I_l \sim c_l \: m_a^{1+l/2}\,, \quad \mathrm{for} \quad m_a\to\infty \,,
\end{eqnarray}
with positive coefficients $c_l$.

Inserting the asymptotic results for $I_l$ and $I_n$
into the bound (\ref{finalbound}), we get ($\lambda$ is a positive constant)
\begin{eqnarray}
0\le I_{+--} \le \lambda\, \lim_{m_a \to \infty}\;
\sum_{l,n=0}^4 H(l,n)\,m^{1+l+n}\, c_l \, c_n \, m_a^{-1-l/2-n/2} =0 \,,
\end{eqnarray}
so that
\begin{eqnarray}
I_{+--} =0  \,,
\end{eqnarray}
which implies Eq.~(\ref{Gamma+to+--}) in the main text.

\vspace{-0.25\baselineskip}

\end{document}